\documentclass[journal=apchd5]{achemso}
\usepackage{graphicx}  
\usepackage{amssymb}   
\usepackage{amsmath}
\usepackage[T1]{fontenc} 
\usepackage[latin1]{inputenc}
\usepackage{lmodern} 
\usepackage{color}

\usepackage{psfrag}

\usepackage{graphicx,xcolor}
\usepackage{amsmath, bbm}
\usepackage{natbib}
\usepackage{hyperref}
\usepackage{epstopdf}
\usepackage{bbm}
\usepackage{soul}

\renewcommand{\imath}[0]{\mathrm{i}}

 \title{Manipulation of quenching in nanoantenna-emitter systems enabled by external detuned cavities: a path to enhance strong-coupling}

\author{Burak Gurlek}
 \affiliation{Max Planck Institute for the Science of Light, Staudtstra{\ss}e 2, D-91058 Erlangen, Germany}
\alsoaffiliation{Friedrich Alexander University Erlangen-Nuremberg, D-91058 Erlangen, Germany}
\author{Vahid Sandoghdar}
\affiliation{Max Planck Institute for the Science of Light, Staudtstra{\ss}e 2, D-91058 Erlangen, Germany}
\alsoaffiliation{Friedrich Alexander University Erlangen-Nuremberg, D-91058 Erlangen, Germany}
\author{Diego Mart\'in-Cano}
\email{diego-martin.cano@mpl.mpg.de }
  \affiliation{Max Planck Institute for the Science of Light, Staudtstra{\ss}e 2, D-91058 Erlangen, Germany}
  
\keywords{Quenching, nanoantennas, plasmonics, microcavities, strong-coupling, single-photon emitters.}
\date{\today}
\begin{tocentry}
 \includegraphics{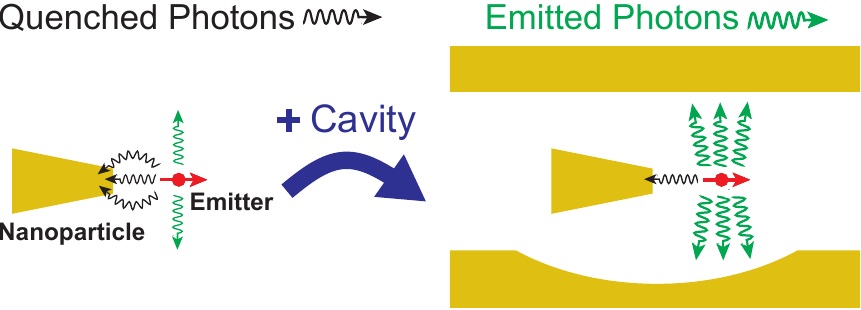}
\end{tocentry}
\begin{document}
\begin{abstract}
We show that a broadband Fabry-Perot microcavity can assist an emitter coupled to an off-resonant plasmonic nanoantenna to inhibit the nonradiative channels that affect the quenching of fluorescence. We identify the interference mechanism that creates the necessary enhanced couplings and bandwidth narrowing of the hybrid resonance and show that it can assist entering into the strong coupling regime. Our results provide new possibilities for improving the efficiency of solid-state emitters and accessing diverse realms of photophysics with hybrid structures that can be fabricated using existing technologies.
\end{abstract}




\maketitle

The excited state of a quantum emitter can decay radiatively via spontaneous emission of photons or nonradiatively in a process called quenching. The interplay between these two decay channels crucially determines the application potentials of solid-state emitters such as organic molecules, semiconductor nanocrystals or color centers \cite{Lounis2005}. While spontaneous emission is known to be enhanced or inhibited by photonic environments \cite{Haroche2006}, the nonradiative decay channel is usually thought to be an intrinsic property of the emitter and its immediate surrounding. 

The best-known modification of radiative rates is the so-called Purcell effect, where a quantum emitter is coupled to a conventional resonator of quality factor $Q$ and mode volume $V$ \cite{Purcell1946}. When the atom-photon interaction rate becomes larger than both the cavity loss rate ($\kappa$) and the atomic coupling rate to other competing modes, one also can reach the strong-coupling regime (SCR) \cite{Haroche2006}, where photonic and atomic excitations are coherently exchanged and hybridized. 

A more recent alternative approach for accessing the Purcell effect or the SCR places the emitter in the near field of plasmonic nanoantennas \cite{Truegler2008, Savasta2010, Agio2012b, Santhosh2016,Chikkaraddy2016, Matsuzaki2017}. However, the close vicinity of the emitter to metals results in dissipation and substantial coupling to higher-order multipolar antenna modes \cite{Ruppin1982,Rogobete2007}, which in turn, causes an increase in the nonradiative rate that is faster than those in the radiative decay \cite{Agio2012b,Ruppin1982}. So far, few nanoantenna configurations \cite{Agio2012b,Santhosh2016,Chikkaraddy2016, Matsuzaki2017} have succeeded in accessing interesting radiative effects in competition with the nonradiative channels.

\begin{figure}
\begin{center}
\includegraphics[width=8.cm]{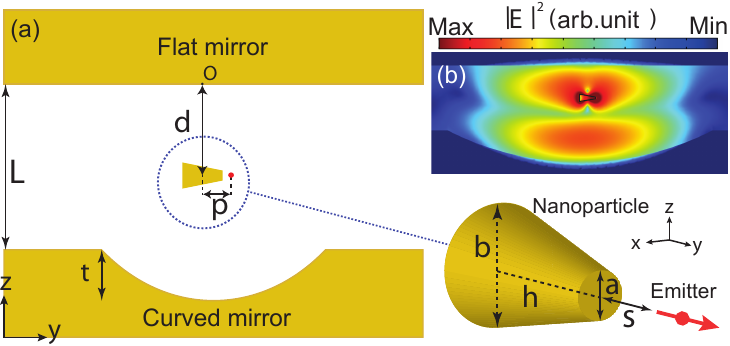}
\end{center}
\caption{(a) Sketch of the FP-nanoantenna hybrid: a dipole emitter close to a plasmonic particle is embedded in a FP cavity. The axially symmetric cavity has a curved mirror radius $R$ and depth $t$ spaced at length $L$. The emitter position is characterized by $p$ and $d$, from the origin $O$.  Inset: Zoom of the emitter placed at distance $s$ from a metallic nanocone of length $h$, and tip and base diameters $a$ and $b$, respectively. (b) Intensity distribution of the detuned FP-nanoantenna mode (see also Fig. \ref{fig2}). \label{fig1}}
\end{figure}

In this Letter, we study the coupling of a quantum emitter to a hybrid structure consisting of a Fabry-Perot (FP) resonator and a plasmonic nanoantenna. Figure \ref{fig1} sketches an example of the proposed device using a gold nanocone antenna \cite{Mohammadi2010}. Hybrid arrangements have recently considered the combination of cavities with plasmonic nanoantennas for achieving Purcell enhancement \cite{Frimmer2012,Dezfouli2016,Doeleman2016} and strong coupling \cite{Xiao2012}. In what follows, we explore regimes where both radiative and nonradiative properties of an emitter are improved if a cavity is hybridized with a strongly detuned nanoantenna.  Importantly, we demonstrate that one can generally counteract and control nonradiative channels from afar using a FP resonator. In contrast to previous works, we also show that an emitter coupled to the hybrid compound resonance can enter the SCR in configurations, where neither the isolated nanoantenna nor the cavity alone would access this regime. Aside from a fundamental interest, these results hold promise for practical applications, where the emitter quantum efficiency plays an important role. 

\begin{figure}[!htb]
\begin{center}
\includegraphics[width=8.3cm,angle=0]{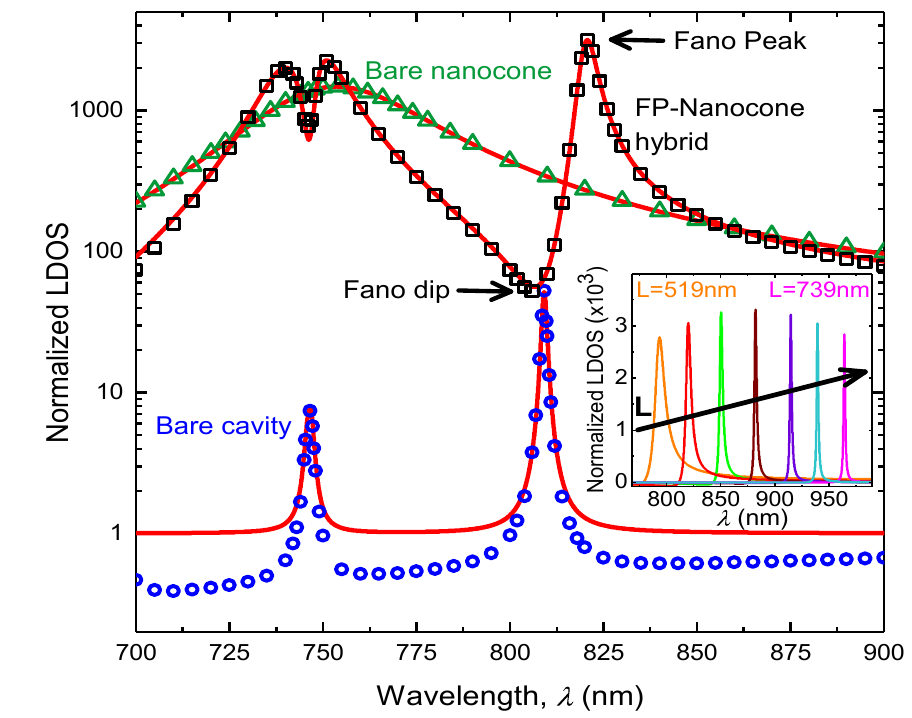}
\end{center}
\caption{LDOS for a FP-nanocone hybrid (black squares), a bare nanocone (green triangles) and a bare cavity (blue circles) versus the emission wavelength (logarithmic vertical scale). All are normalized to the LDOS in free space. The emitter lies at a $s=10$~nm from a nanocone near the cavity center ($p=150$~nm, $d=280$~nm, c.f. parameter definitions in Fig.~\ref{fig1}) with its dipole moment along the y-direction. The nanocone has a length, tip and base of $h=140$~nm, $a=20$~nm and $b=60$~nm, respectively. The cavity parameters ($R=2.5$~$\mu$m, $t=275$~nm and $L=559$~nm) give access to the second $\mathrm{TEM}_{00}$ mode ($\lambda \sim 809$~nm) and a hybrid $\mathrm{TEM}_{01*}$ mode ($\lambda \sim 746$~nm). The dielectric functions for the nanocone and the mirrors were obtained from a Drude-Lorentz fit to experimental data on gold \cite{Mohammadi2010}. The red lines correspond to QNM calculations \cite{Bai2013}. Inset: Normalized LDOS for the detuned hybrid QNM contributions at different mirror lengths $L$. \label{fig2}}
\end{figure}

To investigate the influence of the hybrid cavity-antenna structure on an emitter, we examine the local density of states (LDOS). This electromagnetic quantity is connected to the imaginary part of the Green's tensor and thus to the power dissipated by a dipole, which we calculate by means of full-wave computations with COMSOL Multiphysics \cite{Bai2013}. Figure~\ref{fig2} displays the normalized LDOS for a broadband FP-nanocone system (black squares) as a function of the emission wavelength $\lambda$ for an emitter that lies at ten nanometers from the nanocone close to the antinode of the FP microcavity [see Fig.~\ref{fig1}(b)]. Highly enhancing nanoantennas, such as nanocones, facilitate reaching the SCR with moderately low-Q cavities as shown below. 
We identify two main regions of enhanced LDOS in Fig.~\ref{fig2}: a double-peaked feature with a very broad linewidth at $\lambda \sim 750$\,nm and a narrower resonance around 820\,nm. 

It is instructive to compare the normalized LDOS to the same configurations of a bare microcavity (blue circles) and an isolated nanocone (green triangles) in Fig.~\ref{fig2}. The outcome indicates that the broadband plasmon modes of the nanoantenna and two transverse cavity modes with narrower linewidths interfere constructively to yield to two general scenarios: a double-peaked structure for resonant modes coupling and a shifted cavity resonance for off-resonant interaction. The latter frequency change at longer wavelengths is attributed to the common cavity red shifts reported for small plasmonic nanoparticles \cite{Kelkar2015}. Notice that the maximum LDOS values ($\sim3\cdot10^3$) for the off-resonant mode has been enhanced by one order of magnitude with respect to the bare cavity mode and by a factor of three with respect to the isolated nanocone over a fairly narrow bandwidth. This enhancement comes as a result of the intermediate values of both the quality factor ($Q_{\mathrm{hyb}}\sim150, Q_{\mathrm{cone}}\sim14$, $Q_{\mathrm{FP}}\sim510$) and the mode volume of the hybrid resonance ($V_{\mathrm{hyb}}\sim 3.6\cdot 10^{-3}\lambda^3 ,V_{\mathrm{cone}}\sim 7.2 \cdot 10^{-4}\lambda^3, V_{\mathrm{FP}}\sim 0.75\lambda^3 $). These features make the detuned hybrid mode very attractive for strong coupling as shown below. Furthermore, the combination of the nanoantenna and cavity modes also leads to dips in the LDOS values. Both the peak and dip result respectively from constructive and destructive interference events known from Fano phenomena \cite{Luk2010} for two resonant systems with antithetic bandwidths ($Q_{\mathrm{cone}}\sim14$, $Q_{\mathrm{FP}}\sim510$). 

An important and attractive aspect of the broadband hybrid cavity is that the enhancement effect can be tuned to different frequencies over a very large spectral range by simply adjusting the cavity length $L$ (see inset in Fig.~\ref{fig2}). In fact, it is remarkable that the LDOS is enhanced to such a degree at over hundred nanometers wavelength detuning from the antenna plasmon resonance, which in this case was set close to 750\,nm. Intuitively, the circulation of the optical energy in the microcavity compensates for the lower plasmonic enhancement of the LDOS at a large detuning. We note that this phenomenon provides a unique and novel means for external and selective manipulation of the emitter coupling to plasmonic antennas.
 
\begin{figure}[!htb]
\includegraphics[width=8.4cm,angle=0]{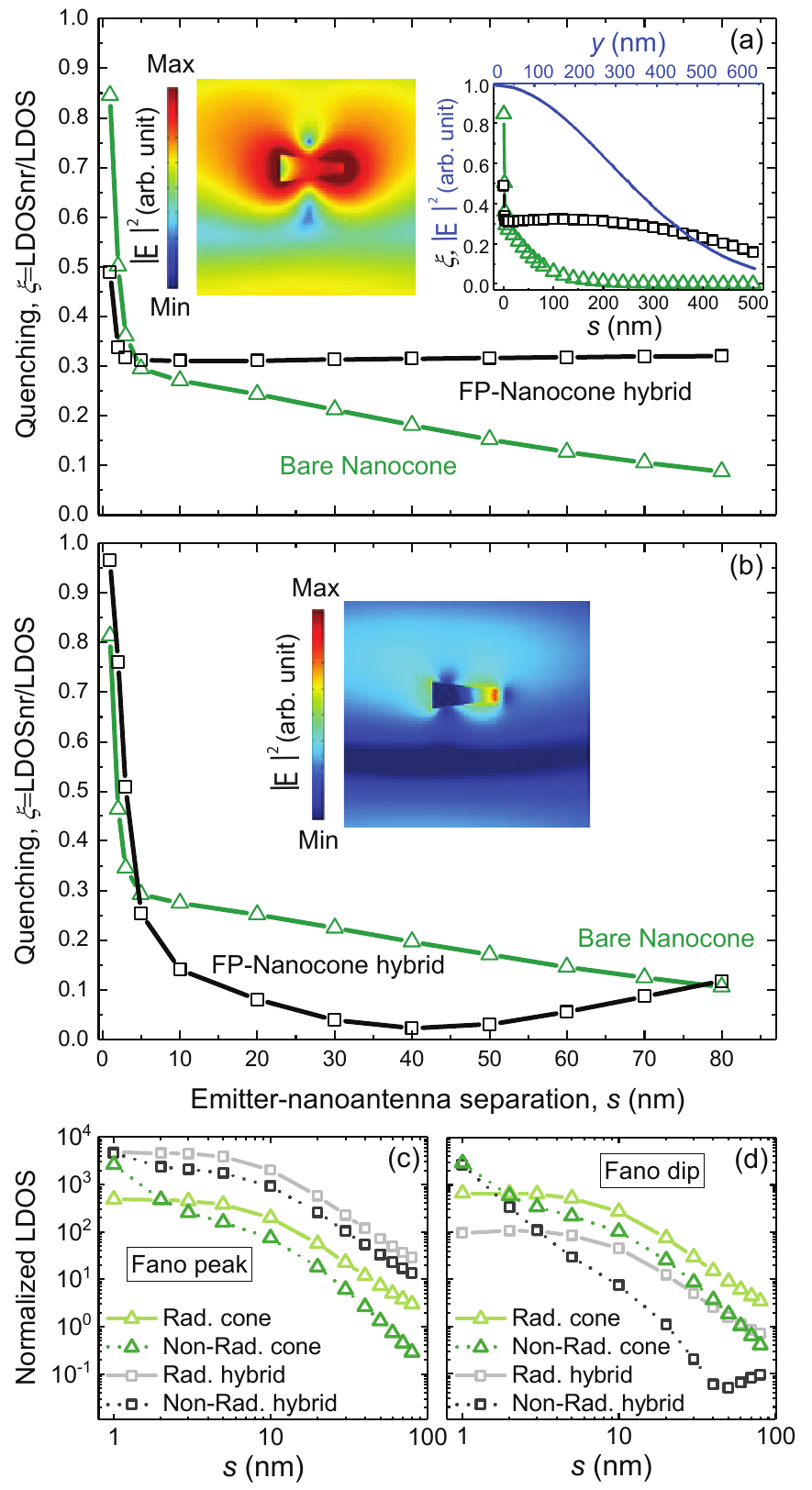}
\caption{Quenching for the detuned FP-nanoantenna mode (black squares) as a function of the separation distance $s$ between the emitter and the nanocone, the latter remaining fixed. Panel (a) considers emission at the Fano peak, $\lambda=820$~nm, and (b) at the Fano dip, $\lambda=806$~nm. The green triangles display the bare nanocone results. The right inset in (a) shows $\xi$ at $s$ far from the nanocone together with the bare cavity intensity along the $y$-position, shown as reference. Left inset, scattered intensity map for a distance $s=40$nm, and analogously in (b). Panels (c) and (d) show both the normalized radiative and nonradiative LDOS [$\mathrm{LDOS_{r}}=(1-\xi)\cdot$LDOS and $\mathrm{LDOS_{nr}}=\xi \cdot$LDOS, respectively] versus $s$, corresponding to panels (a) and (b), respectively. The other configuration parameters remain as in Fig.~\ref{fig2}.\label{fig3}.}
\end{figure}

To obtain a deeper insight into the different participating resonant modes and to evaluate semianalytical expressions of the Green's tensor, we also used the quasinormal mode (QNM) approach that is based on a modal expansion and the Lorentz-reciprocity theorem \cite{Sauvan2013}. This enables to determine the Purcell factor for single QNMs \cite{Sauvan2013} that is valid for any lossy resonator (see implementation in ref \citep{Bai2013}) and it is derived from the Green's tensor with component $c^2 E E/2\omega(\omega-\omega_r-\imath\kappa/2)$ \cite{Collin1990}, where $E$ denotes the normalized field parallel to the orientation of the dipole at its position \cite{Bai2013}, $\omega_r$ is the QNM resonance real frequency, and  $\kappa=\omega_r/Q$ denotes its fullwidth at half-maximum. The red lines in Fig.~\ref{fig2} represent the contribution from several QNMs, showing an excellent agreement for the hybrid full-wave response, whereas additional non-resonant modes would be necessary for describing the suppressed LDOS values with respect to free space (cf. values below 1 for the blue circles). The double-peaked LDOS arises as a result of the interference of two nearly resonant QNMs, consisting of a plasmonic-like mode and an FP-like one with positive and negative values, respectively. Negative contributions are common features of nearly resonant QNMs \cite{Sauvan2013}, with the total sum remaining positive and thus physical (cf. red line). On the other hand, the detuned peak is mainly described by a single FP-nanoantenna QNM [cf. its intensity distribution in Fig.~\ref{fig1}(b)], whereas the broader off-resonant QNMs contribute to the destructive interference dip. 

A severe general limitation of plasmonic nanoantennas concerns quenching of emission at very small distances caused by nonradiative channels \cite{Rogobete2007,Delga2014}. To study the quenching behavior of the detuned hybrid mode, we calculated the fraction of the LDOS that is dissipated ($\mathrm{LDOS}_{\mathrm{nr}}$) in the metallic nanostructure given by $\xi=\mathrm{LDOS}_{\mathrm{nr}}/\mathrm{LDOS}$ as a direct measure for quenching, where LDOS$_{\mathrm{nr}}\propto\int \mathrm{Im}[\epsilon ]|E|^2\mathrm{d}\mathbf{r} $, and $\mathrm{Im}[\epsilon]$ is the imaginary part of the permittivity of the metal. 
Figure~\ref{fig3}(a)-(b) displays $\xi$ versus the antenna-emitter distance $s$ at the Fano peak (a) and dip (b) of the far-detuned hybrid mode shown in Fig.~\ref{fig2}, while Fig.~\ref{fig3}(c)-(d) presents both the radiative and nonradiative components of the LDOS, together with the bare nanocone results for comparison. 

At the Fano peak [cf Fig.~\ref{fig3}(a)], we find that  $\xi$ is stronger than the case of a bare nanocone for $s>5$\,nm and has a nearly constant value. This anomalous trend is in marked contrast to the quenching behavior for a bare nanoantenna (green triangles), which commonly increases in a monotonous fashion for smaller $s$\,\cite{Rogobete2007}. We attribute these findings to the concentration of the field in the metallic structure [see left inset of Fig.~\ref{fig3}(a)] caused by the constructive interference of the nanoantenna and cavity modes, thus, resulting in an enhanced absorption. The right inset in Fig.~\ref{fig3}(a) shows that the circulation of the optical energy in the microcavity can extend this effect even to separations comparable to a FWHM of the cavity mode profile (blue curve), where quenching by a bare nanocone becomes negligible. We note that, nevertheless, the structure keeps an overall good radiation efficiency of $1-\xi\sim 68\%$ over a large spatial range while the LDOS is enhanced by ten times with respect to the bare cone [see Fig.~\ref{fig3}(c)]. This radiative emission can be mostly collected by a FP mode with an overlap of about 81\% [see Fig. \ref{fig1}(b)].

Another impressive phenomenon occurs at the Fano dip, as presented in Fig.~\ref{fig3}(b). Here, $\xi$ acquires lower values than that of the bare nanocone case, exhibiting \textit{suppression} of quenching. The inset in Fig.~\ref{fig3}(b) illustrates that in this case, the destructive interference of the plasmonic and cavity modes lead to an intensity minimum inside the nanocone. This effect makes the emitter highly efficient over a large distance range, e.g. $1-\xi\sim 98\%$ at $s=40$ nm. Figure~\ref{fig3}(d) shows in more detail the involved competition between the radiative and nonradiative rates of the hybrid structure at different emitter-antenna separations compared to the bare nanoantenna case. The balance between emission enhancements and quenching can be adjusted smoothly between the Fano dip and peak by varying the FP cavity length or by changing the emitter-antenna separation. For example, the general quenching behaviors are reversed for $s<5$ nm due to the cavity radiative competition against quasistatic contributions \cite{Gonzalez-Tudela2014,Delga2014}. These above-described cavity modifications are general and thus observable for different nanoparticles (see more examples in Supporting Information). 
 
\begin{figure}[!htb]
\includegraphics[width=8.4cm,angle=0]{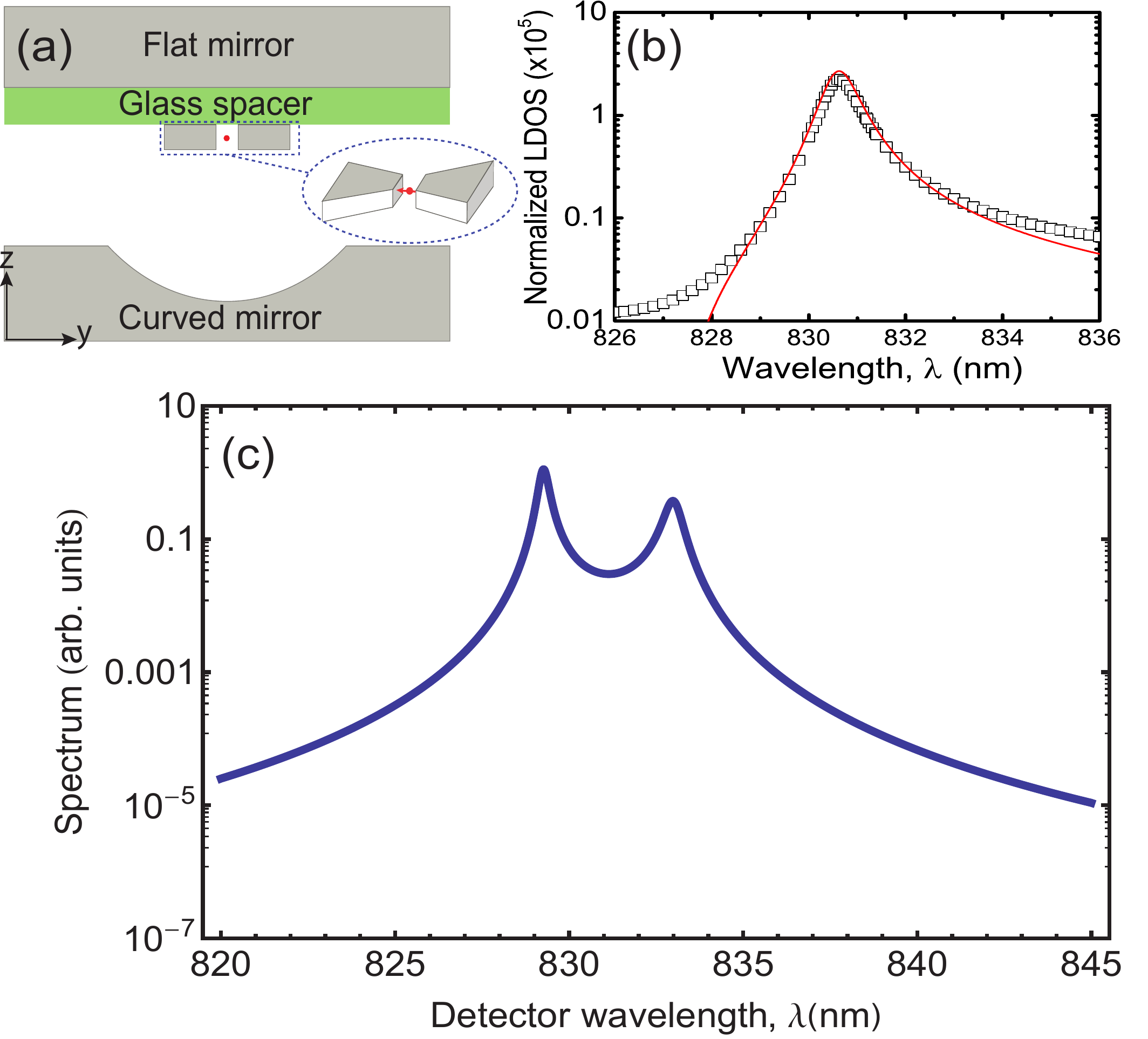}
\caption{(a) FP-bowtie hybrid sketch. The flat cavity mirror has a glass dielectric spacer of $120$~nm and $L=504$~nm, whereas the other parameters remain as in Fig.\,\ref{fig1}. The emitter lies at the center of a bowtie antenna ($p=0$~nm, $d=135$~nm) with a $2$~nm gap and $30$~nm thickness remaining on top of the spacer. The single particles (trapezoidal) have a length of $50$~nm with tip and base sizes of $15$~nm and $40$~nm, respectively. We used the permittivity of silver at $T=4$~K from the Drude-Lorentz model described in Ref.\,\cite{Hummer2013} for the nanoparticles and mirrors. The hybrid-detuned resonance frequency is $361$~THz. (b) Normalized LDOS (black squares) and single QNM calculation (red line) show the hybrid detuned mode for the configuration described in (a). (c) Measurable spectrum of fluorescence for an emitter of linewidth 25 MHz resonant with the detuned-FP-bowtie hybrid mode (blue line) versus detector wavelength.} \label{fig4}
\end{figure}

We now present an example of an antenna-FP geometry that shows an amelioration of the nonradiative channels and brings a single quantum emitter to the SCR. Here, we consider a bowtie antenna \cite{Kinkhabwala2009,Santhosh2016} compatible with fabrication on a flat cavity mirror [see Fig.~\ref{fig4}(a)] and an organic molecule with a typical natural linewidth of $\gamma/2 \pi = 25$ MHz at cryogenic temperatures. We chose the antenna parameters (see caption of Fig.~\ref{fig4}) to place its resonance at $\lambda_r=690$~nm and tune a moderately low-Q cavity resonance ($Q$=3400) at $\lambda_r=826$~nm in order to obtain an enhancing hybrid mode as discussed in Fig.~\ref{fig2}. Figure~\ref{fig4}(b) displays the LDOS enhancement on the Fano peak of the resulting hybrid mode at $\lambda_r=831$~nm, whereas the red line shows the excellent single QNM approximation near resonance that relates directly to the Purcell factor \cite{Sauvan2013}, reaching $F= 2.7\times10^5$.

Applying the macroscopic quantum electrodynamics and the Green's tensor formalism \cite{Vogel2006,Delga2014}, we evaluate the resonance fluorescence spectrum for the composite structure within the single QNM approximation and a two-level emitter (see derivation in ref \citep{Delga2014}). Figure~\ref{fig4}(c) shows the spectrum outcome, which reveals a large peak splitting characteristic of strong coupling. Using a single Lorentzian model \cite{Vogel2006}, provided within the single QNM approximation, we can estimate and understand the coupling rate $g$ and the condition $g=\sqrt{F \gamma \kappa}/2>\kappa/4$ for entering in SCRs in terms of $F$, $\kappa$ and the splitting 2$g$ in the spectrum. In the case of the bare cavity, $Q=3400$, $V=0.44\lambda^3$ and $\kappa=6.7\times 10^{11}~\mathrm{rad}\cdot\mathrm{s}^{-1}$ lead to $F= 587$ and $4g/\kappa=2\sqrt{F\gamma/\kappa}=0.7<1$, i.e. outside the SCR. Similarly, in the case of a bare nanoantenna, the very small mode volume of $4\cdot10^{-5} \lambda^3$, $Q=20$ and $\kappa=1.4\times 10^{14}~\mathrm{rad}\cdot\mathrm{s}^{-1}$ lead to $F=39000$ while $4g/\kappa=0.4$ remains well below unity (see their respective single-peak spectra in Supp. Information). It is the cooperative-enhanced ratio of large $F$ and fairly low $\kappa$ of the detuned FP hybrid mode ($F= 2.7\times10^5$, $V=9.6 \cdot 10^{-4} \lambda_r^3$, $\kappa=2.3\times 10^{12}~\mathrm{rad}\cdot\mathrm{s}^{-1}$) that allows surpassing the strong coupling threshold to $4g/\kappa=8.6$, while achieving a high radiative efficiency $1-\xi=95\%$ enabled by the cavity. This suppression of nonradiative channels occurs precisely at the extreme close distances favorable for the SCR with bare nanoantennas (see details in Supp. Information), showing a potential route to ameliorate the significant nonradiative emission in plasmonic systems that hinders its photonic detection. In addition, we note that $F$ reported by previous hybrid cavity-nanoantenna structures \cite{Xiao2012,Dezfouli2016,Doeleman2016} are about two orders of magnitude lower. 

The features of detuned FP-antenna hybrids studied in this work are general and potentially observable in previously proposed hybrid systems \cite{deAngelis2008,Ameling2010,Benson2011,Chamanzar2011,Vazquez2014,Bahramipanah2015} by including single emitters \cite{Frimmer2012,Xiao2012,Dezfouli2016,Doeleman2016}, with different cavity geometries \cite{Benson2011,Frimmer2012,Xiao2012,Hummer2013,Dezfouli2016,Doeleman2016,Kelkar2015,Wang2016} and nanoantenna structures \cite{Santhosh2016,Chikkaraddy2016}. These open new avenues in diverse research areas such as sensing, surface-enhanced Raman scattering, solid-state spectroscopy~\cite{Benson2011,Orrit2014}, and quantum optics~\cite{Tame2013}. The resulting Fano resonances at tunable frequencies can be used to improve the emission efficiency of solid-state emitters such as organic molecules or color centers by selective enhancement of their zero-phonon transitions \cite{Grange2015,Wang2016}. Important limits to the coupling in these hybrid systems are set by the antenna scatterings and cavity losses \cite{Doeleman2016} that have no general analytical solutions. This calls for an extensive research for optimizing the suppression of quenching on metallic nanoantennas with ultimate radiative enhancements, e.g. in atomic enhanced nanoparticles \cite{Barbry2015}. The strong radiative enhancement also ushers in new studies of single-emitter coherent interactions at ambient temperatures, where $g=\sqrt{F \gamma \kappa}/2>\kappa/4,\gamma^*/2$. Here, $\gamma^*$ accounts for the pure dephasing rate \cite{Cui2006} caused by phononic excitations, which typically ranges from $10^3\gamma$ to $10^5 \gamma$ for solid-state emitters \cite{Grange2015}. In contrast to inefficient high-Q cavity approaches, where the emitter linewidths exceed $\kappa$ by several orders of magnitude \cite{Grange2015}, the high coupling rates ($g/2\pi\sim 0.1-1$~THz) and large bandwidths ($\kappa/2\pi \sim 0.1-1$~THz) of the hybrid structure could then potentially bring diverse solid-state emitters at room temperatures into the SCR (e.g. for a SiV \cite{Grange2015}, $\gamma/2\pi\sim 300$ MHz and $\gamma^*/2\pi \sim 10^{3}\gamma/2\pi\sim 1$~THz, or for a molecule, $\gamma/2\pi\sim 25$ MHz and $\gamma^*/2\pi \sim 10^{5}\gamma/2\pi\sim10$~THz). This order of magnitude estimation points to the accessibility of interesting phenomena with strong coupling as reflected in diverse reports, e.g. strong single-photon nonlinearities~\cite{Imamoglu1997}, nonclassical processing~\cite{Cirac1997,Cirac1995}, Bose-Einstein condensation~\cite{Kasprzak2006}, unobserved optomechanical effects~\cite{Aspelmeyer2014}, and rare chemical reactions \cite{Hutchison2012,Galego2016}. Another prominent possibility is to tune the near-field quenching and enhancement caused by plasmonic and to extend these interactions to longer length scales. This opens important doors for coupling many emitters to the same hybrid mode, facilitating investigations of collective effects such as nonclassical correlations \cite{Carmichael1991} with enhanced dipole-dipole interactions \cite{Haakh2015} and in collective strong coupling phenomena \cite{Gonzalez2013} that take place on femtosecond scales at room temperature \cite{Santhosh2016,Chikkaraddy2016}. Here, the tunability of cavity-antenna systems can be particularly useful for modifying \textit{in-situ} these collective interactions  \cite{Hutchison2012,Gonzalez2013,Delga2014,Torma2015,Galego2016}. 
\begin{acknowledgement}
This work was supported by the Max Planck Society and the European
Research Council (Advanced Grant SINGLEION). V. acknowledges the support from the Alexander von Humboldt
Professorship. The authors thank Claudiu Genes for reading and providing constructive comments of the manuscript.
\end{acknowledgement}
\begin{suppinfo}
Cavity modification of quenching and LDOS for a nanosphere. Quenching suppresion in state-of-the art nanoantennas at small separations. Spectra for bowtie, bare cavity and bowtie-cavity hybrid.
\end{suppinfo}

\end{document}